\documentclass[12pt,preprint]{aastex}
\usepackage{graphicx}

\newcommand{\nod}{\nodata}
\newcommand{\src}{4U~1705$-$440}
\newcommand{\ks}{\ensuremath{K_{\rm s}}}
\newcommand{\chandra}{\textit{Chandra}}

\begin{document}

\title{A variable near-infrared counterpart to the neutron-star
low-mass X-ray binary \src\footnote{This paper includes data
gathered with the 6.5 meter Magellan Baade Telescope located at Las
Campanas Observatory, Chile, and the 4 meter Blanco Telescope located
at CTIO, Chile.}}

\author{Jeroen Homan, David L.\ Kaplan\footnote{MIT Pappalardo and Hubble Fellow}}

\affil{MIT Kavli Institute for Astrophysics and Space
Research, 70 Vassar Street, Cambridge, MA 02139 \\
jeroen@space.mit.edu,dlk@space.mit.edu}

\author{Maureen van den Berg}

\affil{Harvard-Smithsonian Center for Astrophysics, Cambridge, MA
02138 \\ maureen@head.cfa.harvard.edu }

\and
\author{Andrew J.\ Young}

\affil{Astrophysics Group, Department of Physics, Bristol University, 
Bristol, BS8 1TL, UK \\ andy.young@bristol.ac.uk }

\begin{abstract}

We report the discovery of a near-infrared (nIR) counterpart to the
persistent neutron-star low-mass X-ray binary \src, at a location consistent 
with its recently determined {\it Chandra} X-ray position. The nIR source is
highly variable, with $K_s$-band magnitudes varying between 15.2 and
17.3 and additional $J$- and $H$-band observations revealing color
variations. A comparison with contemporaneous X-ray monitoring observations shows that the nIR brightness correlates well with X-ray flux and X-ray
spectral state. We also find possible indications for a change
in the slope of the nIR/X-ray flux relation between different X-ray states.
We discuss and test various proposed mechanisms for the nIR emission
from  neutron-star low-mass X-ray binaries and conclude that the nIR
emission in \src\ is most likely dominated by X-ray heating of the
outer accretion disk and the secondary star.

\end{abstract}


\section{Introduction}

\src\ is a neutron-star low-mass X-ray binary (NSXB) that was
discovered in the early 1970's with the Uhuru satellite
\citep{gimugu1972}. The X-ray spectral and variability properties place
it among the class of the so-called atoll sources \citep{hava1989}, 
persistent NSXBs that  accrete between  a few and a few tens of
percents of the Eddington luminosity. The source shows type-I
(thermonuclear) X-ray bursts \citep{stlafr1985,gohala1989}, strong
rapid variability \citep{olbagi2003}, including kHz quasi-periodic
oscillations \citep{fovaka1998}, and some of the strongest long-term X-ray
variations seen in persistent NSXBs \citep{repaky2003}. These long-term variations 
correspond to X-ray spectral-state transitions, as was
shown by \citet{baol2002}. 

Despite the fact that \src\ has been known for more than 35 years, no
optical, infrared, or radio counterparts have been identified. Given
the pronounced X-ray state transitions and large luminosity changes in
\src, such counterparts are expected to be variable as well, which
should in principle facilitate identification efforts. The main reason
that counterparts have not been found is that, until recently, no
accurate X-ray source position was available, impeding identification efforts in the crowded Galactic-bulge field. This changed with a {\it
Chandra} observation by \citet{diiame2005}, which yielded source
coordinates with an absolute uncertainty of $\sim$0\farcs5. As noted by
\citet{diiame2005}, these new coordinates differ significantly from
previously reported ones, by $\sim$0\farcm15.


We have conducted a search for the near-infrared (nIR) counterpart of
\src, taking advantage of the improved {\it Chandra} position. The
main benefit of searching in the nIR in comparison to the optical is
the much lower extinction; based on $N_H$ estimates from X-ray spectra
\citep{pisadi2007} we expect $A_V\approx10.8$, but only $A_K\approx1.3$.
Moreover, as was recently shown by \citet{rufehy2006,rufejo2007}, the
nIR can provide important constraints on the emission properties of
the jet and accretion disk in NSXBs. In particular, for atoll sources
\citet{rufejo2007}  found that above an X-ray luminosity of $10^{36}$
erg\,s$^{-1}$ the nIR is dominated by jet emission, while below these
luminosities X-ray reprocessing in the accretion disk dominates
(although thermal emission from the disk cannot be ruled out).

In this paper we report the discovery of a variable nIR counterpart to
\src. In \S\ref{sec:obs} we present our nIR observations along with a
set of quasi-simultaneous X-ray observations. The identification of
the nIR counterpart and a study of nIR/X-ray correlations are presented in
\S\ref{sec:results}. Finally, in  \S\ref{sec:discuss}, we discuss our
results in the framework of several proposed mechanisms for nIR
emission in NSXBs.

\section{Observations \& Data Reduction}\label{sec:obs}

\subsection{near-Infrared}

We observed the field of \src\ a number of times in the nIR between 2006 May 16 and 2007 September 22. Our
primary observations were with the 6.5 m Magellan I (Baade) telescope
and Persson's Auxiliary Nasmyth Infrared Camera
\citep[PANIC;][]{mapemu2004}, which contains a $1024\times1024$
HAWAII~I HgCdTe detector with a $0\farcs125$ plate scale for a
$128\arcsec\times128\arcsec$ field of view.  We also obtained two
observations with the 4 m Blanco Telescope at the Cerro Tololo
Inter-American Observatory (CTIO), using the Infrared Side Port Imager
(ISPI).  ISPI has a $2048\times2048$ HgCdTe HAWAII-2 array with
$0\farcs3$ plate scale, for a $10\farcm25\times10\farcm25$ field of
view, although to match the PANIC data and to avoid dealing with
geometric distortions we only used the central $512\times512$ pixels. 
The majority of the data were taken with a \ks\ filter, but we did
obtain some observations with the $J$ and $H$ filters.  A log of the
observations is given in Table~\ref{tab:obs}.  There we list the total
exposure times for each observation, although the data actually
consist of a number of short (10--20~s for \ks, and up to 48~s for the
other filters) exposures taken at positions dithered by up to
$10\arcsec$.

We reduced the data using custom routines in \texttt{PyRAF}.  First we
linearized the data using the coefficients given on the
PANIC\footnote{\url{http://www.lco.cl/lco/telescopes-information/magellan/instruments-1/panic/panic-documentation/panic-manual/panic-observers-s-manual/hgcdte-detector/linearity-saturation}.}
and
ISPI\footnote{\url{http://www.ctio.noao.edu/instruments/ir\_instruments/ispi/
New/UsersGuide/linearity.html}.}
web sites.  We then subtracted dark frames, divided by a flat field
(using twilight flats for PANIC and dome flats for ISPI), and then
produced a sky frame for subtraction by averaging the four exposures
on either side of a given exposure. We then added the sky-subtracted
exposures together, identified all the stars, and produced masks for
the stars to improve the sky frames in a second round of sky
subtraction.  The coaddition of the individual exposures shifted
the images of all the bands from a given observation to the same position.

For a reference observation, we used Epoch~1 from Table~\ref{tab:obs}
(2006~May~16).  We determined the astrometry using 64 2MASS
\citep{skcust2006} stars that were well isolated and not very
saturated, fitting for a plate scale, rotation, and zero point.  The
solution for the \ks\ image had an rms of $0\farcs07$ in each
coordinate, and we transferred that solution to the other bands.  For
photometry, we used 34 2MASS stars that were unsaturated in all bands,
well detected in 2MASS, and well isolated in all the data.  We
determined a zero point for each band (not including any color terms),
and got zero-point uncertainties of 0.09~mag.

All of the other data were corrected to the astrometry and
photometry of Epoch~1.  A list of star
positions and magnitudes for 820 stars that were well-detected was produced by running \texttt{sextractor}
\citep{bear1996} on the stacked images of each band of Epoch~1.  We
then ran \texttt{sextractor} to measure each of those stars in all of
the other PANIC images.  To compute the astrometric solution we used about 450 stars in
each \ks\ exposure and got rms residuals of $\approx 0\farcs03$ in
each coordinate.  We then transferred the solutions to the
$J$ and $H$ bands.  The photometric referencing was done using 16
stars within $40\arcsec$ and $\pm2$~mag of \src\ (based on the
2006~May~16 observation; see below), shown in Figure~\ref{fig:jhk}. 
We selected these stars by searching through the 820 initially
identified stars for those with the lowest variation between all of
the exposures, after correcting for different zero points.  These stars
had rms variations of $<0.05$~mag over all the \ks\ observations.

\subsection{X-rays}

During part of the period in which the nIR observations of \src\ were done,
the source was monitored in X-rays with a number of
instruments. For this paper we made use of publicly available data
from the All-Sky Monitor\footnote{http://xte.mit.edu/ASM\_lc.html}
\citep[ASM;][]{lebrcu1996} onboard the {\it Rossi X-ray Timing
Explorer} ({\it RXTE}) and the {\it Swift} Burst Alert
Telescope\footnote{http://heasarc.gsfc.nasa.gov/docs/swift/results/transients/}
\citep[BAT;][]{babacu2005}. The ASM and BAT typically observed the
source a few times per day in the 2--10 keV and 15--150 keV bands,
respectively. In the period between 2007 Aug
11 and 2007 Sep 11 no BAT data were available due to
attitude control problems on the {\it Swift} spacecraft


Additional X-ray data from pointed observations with the
Proportional Counter Array \citep[PCA;][]{jamara2006} and High Energy
X-ray Timing Experiment \citep[HEXTE;][]{roblgr1998} onboard {\it
RXTE} were analyzed. These instruments provide coverage in the 2--60 keV and 15--250
keV bands, respectively.  Starting from 2007 June 26 we observed \src\
on average once every four days. The PCA and HEXTE
data were analyzed with  FTOOLS V6.3. We obtained background-corrected
count rates from the PCA and HEXTE in energy bands that closely
match the energy bands of the ASM and BAT: 3--10 keV for the PCA and
15--150 keV for HEXTE. 

The PCA and HEXTE data were also used for a spectral analysis. For the
data reduction and modeling of the spectra we followed the data-reduction and analysis steps
taken by \citet{lireho2007}. Spectral fits were made in the 3.5--150
keV band, with the $N_H$ fixed to $1.9\times10^{22}$ atoms\,cm$^{-2}$
\citep{pisadi2007}.

\section{Results}\label{sec:results}

\subsection{Near-Infrared Counterpart Identification}

Examining our Epoch~1 data, we find a single bright object within the
\chandra\ error circle (J2000): $\alpha=17^{\rm h}08^{\rm m}54\fs47$,
$\delta=-44\degr06\arcmin07\farcs4$, with an uncertainty (90\%) of $0\farcs6$
determined by the \chandra\ absolute pointing uncertainty (see
Fig.~\ref{fig:jhk}).  This nIR source, at $\alpha=17^{\rm h}08^{\rm
m}54\fs48$, $\delta=-44\degr06\arcmin07\farcs4$, is well within the
\chandra\ error circle, and  is consistent with an unresolved point
source in all of the data.  In the Epoch~1 observations, this source
has $\ks=15.23\pm0.02$, $H=16.03\pm0.02$, and $J=17.14\pm0.02$, where
the errors only reflect the measurement uncertainties.


When comparing all observations, it becomes apparent that the source
varies considerably. These variations are not caused by the different
seeing conditions. In Figure \ref{fig:phot} we plot all our
photometric data, with the photometry listed in Table~\ref{tab:phot}.
Uncertainties of $\sim$0.05 mag arising from our flat-fielding and
sky-subtraction procedures were not included in the Table, but were
taken into account in all figures. We include the two best comparison
stars in  Figure \ref{fig:phot}, which have rms brightness variations
of $<0.06$~mag.  In the $K_s$-band, in which we have the most
observations, the counterpart varies by as much as 2~mag. 
Similar variations are seen in the other bands, although there are
considerable color variations; $H-K_s$ varies by $\sim$0.4 and $J-K_s$
by $\sim$0.6. These color changes appear to be uncorrelated to
changes in nIR brightness. In Figure \ref{fig:sed} we plot a few
representative nIR spectral energy distributions. The data for the
spectral energy distributions were de-reddened assuming an $N_H$ of
1.9$\times10^{22}$ atoms\,cm$^{-2}$ \citep{pisadi2007}, which was
converted to various reddening factors at particular wavelengths, following \citet{prsc1995},
\citet{caclma1989}, and \citet{dusabi2002}: $A_{K_s}=1.27$, $A_H=2.04$, and $A_J=3.03$. A comparison with the slopes of disk and jet
spectra, which are also shown in Figure \ref{fig:sed}, seems to rule
out significant optically thin jet emission in the nIR (see also \S
\ref{sec:discuss}) .

\subsection{Near-Infrared/X-ray correlations}\label{sec:correlations}

The X-ray light curves from the ASM (one-day averages), the BAT
(four-day averages), and PCA/HEXTE (single observation averages) are
shown in Figure \ref{fig:curve}. The times of the nIR observations are
indicated in the panel with ASM data and the corresponding nIR epoch
numbers are given on top. The X-ray state-transition cycles can
clearly be seen, with the 15--150 keV count rates increasing during
periods of low 2--10 keV flux. The nIR properties of \src\ were
studied as a function of X-ray flux and as a function of position in
X-ray color diagrams. 

\subsubsection{X-ray fluxes}\label{sec:fluxes}

For the X-ray flux study we only made use of PCA and HEXTE data, since
the ASM and BAT data are not suitable for spectral fitting. We only
used PCA/HEXTE data that were taken within two days of our nIR
observations; this resulted in 10 observations, two of which were
combined (for nIR Epoch 10). The PCA and HEXTE spectra were fit in
XSPEC V11.3 \citep{ar1996} with simple phenomenological models
\citep{lireho2007}. For  spectrally-hard observations this model
consisted of a black body, to represent the boundary layer between the
disk and neutron star, and a  broken power law, for the non-thermal
emission. For spectrally-soft spectra an additional  disk black body
was needed, to fit the accretion disk component. The models also
included an  absorption component that was fixed at 1.9$\times10^{22}$
atoms\,cm$^{-2}$. These models  resulted in good fits for all our
observations. Unabsorbed 3--100 keV fluxes were extracted and in
Figure \ref{fig:versus} we plot the $K_s$-band flux density versus the
3--100 keV X-ray flux; error bars for the X-ray fluxes were set at
5\%. A clear, positive correlation can be seen; fits with a power law
resulted in a nIR/X-ray slope of 0.66$\pm$0.02 (see dashed line in
Fig.\ \ref{fig:versus}). Separating the data based on relative
strength of the thermal (blackbody and disk black body) and
non-thermal (power-law) components in the  X-ray spectrum, suggests
that the nIR/X-ray flux relation is considerably steeper for non-thermal
dominated (3-100 keV flux contribution $>75\%$) X-ray spectra
(1.33$\pm$0.11; open squares) than in the thermal dominated ($>85\%$)
ones (0.37$\pm$0.03; filled circles), however, we note that the number
of points in both selections is small and that the non-thermal data
points span a very narrow range in X-ray flux. We also point out that
the low energy boundary of 3 keV means that a large fraction of the flux falls outside our flux  band, especially for the soft
spectra.

\subsubsection{X-ray color diagrams}


Spectral states of NSXBs are often
studied in terms of color-color or color-intensity diagrams. Since the
count rates from the ASM, BAT, and HEXTE  were often too low to
produce reliable colors from them, our diagrams were created from PCA data 
only. The PCA data were supplemented by publicly available data of
\src\ from {\it RXTE} observation cycles 9 and 10 (February 2004 until
November 2006), allowing us to put our own observations in a better
context. The color-color and color-intensity diagrams are shown in
Figure  \ref{fig:cd}, in which we also identify the spectrally-hard
(``extreme island"), transitional (``island"), and soft (``banana")
states \citep{lireho2007,va2006}. These spectral states appear as
well-separated branches in the color-color diagram. Since only nine of
our nIR observations could be matched with PCA data  we explored an
alternative way to identify spectral states that could also be
applied to the ASM, BAT, and HEXTE data of \src. As a solution we
simply plotted count rates from a low energy band ($\sim$2--10 keV)
versus count rates from a high energy band ($\sim$15-150 keV). This
was done for the  ASM and BAT four-day averaged count rates and
separately for  our PCA and HEXTE data. To allow comparison between
the ASM/BAT and PCA/HEXTE count-rate diagrams, the count rates of the
various instruments were renormalized. First, the ASM and BAT count
rates were scaled by their maximum four-day averaged values (from
at least 3$\sigma$ detections) from the period after the start of the
BAT coverage (25.6 counts s$^{-1}$ and 2.1$\times10^{-2}$ counts
s$^{-1}$, respectively). The scaling factors for the PCA and HEXTE
light curves were determined visually by finding the best overlap with
the renormalized ASM and BAT light curves, resulting in scaling factors of
869 counts s$^{-1}$ for the PCA and 17.4 counts s$^{-1}$ for HEXTE.
The scaled ASM/BAT and PCA/HEXTE count rate diagrams are shown in
Figure \ref{fig:iid}. Both the ASM/BAT and PCA/HEXTE diagrams reveal
similar two branched structures. The symbols/colors for the PCA/HEXTE
panel are the same as for Figure \ref{fig:cd} and a comparison with
that figure shows that the left-side branch corresponds to the
spectrally-hard (``extreme island") state and the right-side branch
corresponds to the soft (``banana") state. The transitional
(``island") state is not well separated in these diagrams, largely
overlapping with the hard state branch. Note that the branches of the
ASM/BAT track are slightly shorter than those from PCA/HEXTE, mainly
because of missing {\it Swift} data in the August/September 2007
period.

To understand how the broadband X-ray flux changed along the tracks in
Figure \ref{fig:iid} we fitted the spectra of all the PCA/HEXTE
observations in that figure (109 in total) using the same model as
above and obtained unabsorbed fluxes in the 0.5--100 keV band, by
extrapolating well below our low-energy boundary. The main reason for
extrapolating down to 0.5 keV was to capture more flux from the
thermal components, although uncertainties are involved in
extrapolating to such low  energies. We  found that on the
spectrally-hard branch the 0.5-100 keV flux correlated well with the
HEXTE count rate and that on the spectrally soft branch the 0.5-100
keV flux correlated well with the PCA count rate. This suggests that
mass-accretion rate increases as the source moves up either branch.
The 0.5-100 keV flux range observed on the hard branch was
1.1--8.5$\times10^{-9}$ erg\,s$^{-1}$\,cm$^{-2}$; on the soft branch
we measured 1.8--24$\times10^{-9}$  erg\,s$^{-1}$\,cm$^{-2}$.
Considering only the ten X-ray observations that were associated with
the nIR observations, we find that the ones on the spectrally-soft
branch all had higher 0.5--100 keV fluxes that the ones on the
spectrally-hard branch.

Using the same PCA observations as for our spectral analysis
(\S\ref{sec:fluxes}), we overplot the nIR data (grey circles) in the
color-color and color-intensity diagrams in Figure \ref{fig:cd}, with
the size of the symbols scaling linearly with the nIR flux density.
Clearly, the brightest nIR observations occurred in the spectrally-soft state, in which the X-ray flux reaches its maximum. We also
selected ASM/BAT data that were closest in time to our nIR
observations, using three- or four-day ASM/BAT averages that were
centered on the times nIR observations. These data are plotted in
Figure \ref{fig:iid}, together with the PCA/HEXTE data (for which we
used the same nIR/X-ray combinations as for Figures \ref{fig:versus}
and \ref{fig:cd}). Clearly, a large part of the full X-ray spectral
range was covered by our nIR observations. The locations of the nIR
epochs that were covered by ASM/BAT as well as PCA/HEXTE observations
were consistent between the two diagrams, except for Epoch 18, for
which we derived a scaled ASM count rate of 0.96(3) counts s$^{-1}$,
but a scaled PCA count rate of 1.13(1) counts s$^{-1}$. As can be seen
from Figure \ref{fig:iid}, the additional ASM/BAT data strengthen our
initial finding from Figure \ref{fig:cd} that nIR fluxes are highest
in the spectrally-soft state.  In Table \ref{tab:phot} we have listed
the spectral branch on which each of the nIR observations occurred;
two observations occurred near the vertex of the two branches and
identification is therefore uncertain. Comparison of Figure
\ref{fig:iid} and Table \ref{tab:phot} shows that all soft-state
observations have higher nIR fluxes than the hard-state observations.
Since this is also true for the X-ray fluxes, the spectral-state dependence of the nIR flux might therefore simply
reflect the nIR dependence on X-ray flux (see also \S
\ref{sec:discuss}).

\section{Discussion and conclusions}\label{sec:discuss}

We have made nIR observations of the field of the NSXB \src\ and
discovered a variable counterpart at the location of the {\it Chandra}
X-ray source. The $K_s$-band magnitude varied between 15.2 and 17.3
and additional $J$- and $H$-band observations revealed color changes
that did not correlate with nIR brightness variations. Using 18 epochs of
nIR observations in combination with X-ray monitoring observations, we
were able to study, for the first time, the nIR changes along an atoll color-color
track. We found that \src\ has a significantly higher nIR flux in the
spectrally-soft (``banana") state than in the spectrally-hard
(``extreme-island") state, although this might simply be the result of
the higher X-ray fluxes (and mass-accretion rates)  in the former
state. The nIR flux density correlated well with X-ray flux, with
possible indications for a steeper dependence in the spectrally-hard
state. 

Comparing the monochromatic nIR luminosity of \src\ (log$(L_{K_s})\sim19.2-20.0$ [7.2 kpc], $\sim19.5-20.3$ [9.6 kpc]; for distance estimates, see below) with that of other NSXBs published in \citet{rufejo2007}, we find values that are very similar to systems with comparable X-ray luminosities. For $F_{3-100\,keV}/F_{K_s}$ we find values of 1.4--4.4$\times10^{3}$. \citet{rufehy2006,rufejo2007} discussed possible mechanisms  for the
nIR emission in NSXBs: optically-thin emission from the inner regions
of a jet outflow close to the neutron star, thermal emission from an
X-ray or viscously-heated disk, or thermal emission from the
companion star. X-ray heating of the secondary was not explicitly considered by \citet{rufehy2006,rufejo2007}, even though it can be a significant source of optical emission, as shown by \citet{zucash2000}, and presumably also of nIR emission. However, since it is difficult to distinguish heating of the disk from heating of the secondary with our data, we will not discuss the latter option separately.

We start with an assessment of the contribution from a non-heated secondary star. The mass donors in low-mass X-ray binaries are typically late-type dwarfs or giants. We compare the absolute $K_s$ magnitudes of \src\ with published $K$ values for K4 and M4 dwarfs, and K5 and M5 giants \citep{bebr1988,be1991,mihe1982,caos1996}, calculated using the distance estimates derived at the end of this section and a correction for absorption that is based on an $N_H$ of 1.9$\times10^{22}$
atoms\,cm$^{-2}$. For dwarfs we find that they are three to seven magnitudes fainter than our lowest observed values. The absolute magnitudes of the giants, on the other hand, are three to four magnitudes brighter than our highest observed values. Note that these comparisons suffer from considerable distance and reddening related uncertainties. Nevertheless, these findings, in combination with the amplitude of the nIR variations, appear to rule out that thermal emission from a non-heated late-type dwarf or giant dominates the nIR emission of \src. Given that the expected nIR emission from giants is higher than what we observe, the secondary in \src\ is most likely a dwarf. This would then imply an orbital period of $\sim$1--10 hrs for a Roche-lobe filling primary, which needs to be tested with follow-up photometry.

To distinguish between the
other possible mechanisms that are listed above, \citet{rufehy2006,rufejo2007} employed nIR/X-ray
luminosity correlations and spectral energy distributions. In Figure \ref{fig:sed} we compare the nIR spectral energy
distributions of \src\ with models for optically-thin jet synchrotron
emission \citep[$\propto\nu^{-0.6}$; average slope taken from][solid
grey line]{fe2006}, the Rayleigh-Jeans tail of a multi-color disk
black-body spectrum \citep[$\propto\nu^{2}$;][dashed line]{frkira1992}
and the $\nu^{1/3}$ slope of a multi-color disk spectrum that is characteristic for 
frequencies higher than the Rayleigh-Jeans part (dotted
line). The frequency of the break between the $\nu^{2}$ and
$\nu^{1/3}$ slopes of a multi-color disk black-body increases with
mass-accretion rate and decreases with the size of the accretion disk.
Slopes for an X-ray-heated accretion disk are not included in Figure
\ref{fig:sed}. Such heating results in an additional bump in the
optical/UV \citep{vrraga1990} and also extends the high-frequency end
of the Rayleigh-Jeans tail to higher frequencies. 
Based on the model
slopes in Figure  \ref{fig:sed}, a significant contribution to the nIR
from optically-thin jet synchrotron emission can be ruled out;
the slope of the nIR spectrum is the opposite of that expected from a
jet and the relatively small changes in the shape of the spectrum also
 rule out significant changes in the nIR contribution from a
jet across the observed nIR luminosity range. Recent {\it Spitzer} observations of another 
atoll source NSXB, 4U
0614+091, by \citet{mitoma2006}, suggest that one has to observe at
lower IR frequencies to detect clear spectral signatures from a jet in
NSXBs. In that source the jet synchrotron emission had a slope of
$-0.57$, indicative of optically-thin emission and close to the $-0.6$ slope we adopted in 
Figure \ref{fig:sed}. The dependence of nIR flux on X-ray spectral state also argues against a jet 
interpretation of the nIR flux. Observations of black-hole LMXBs \citep{hobuma2005,buba2004} have 
shown that the nIR contribution from a jet changes significantly during transitions between 
spectrally-hard and soft X-ray states, with the nIR contribution in hard states being higher. 
In \src\ we observe the opposite behavior, suggesting that the nIR flux is dominated by other components. 

The slopes of the spectral energy distributions in Figure \ref{fig:sed} are consistent with those of 
multi-color disk black body models, being closer the Rayleigh-Jeans 
tail on the low-frequency
end and closer to the flatter part of a disk spectrum on the high-frequency end. We note that in recent literature on \src\ $N_H$ values in the range of 1.2--2.4$\times10^{22}$ atoms\,cm$^{-2}$ can be found (we adopted 1.9$\times10^{22}$ atoms\,cm$^{-2}$). The low value of this range brings the slope closer to that of the flat part of a disk spectrum, but still far from the jet slope, whereas the high value steepens the data to bring them closer to the slope of the Rayleigh-Jeans 
tail.  Based on the spectral energy distributions we
are not able to distinguish between X-ray-heated or viscously-heated
disk spectra. NIR/X-ray flux correlations could in principle be used to
shed more light this. \citet{rufejo2007} used such correlations for a
large ensemble of NSXBs that covered more than five decades in X-ray
luminosity. Although our X-ray observations of \src\ only span about a
decade in X-ray luminosity, a clear correlation between X-ray and nIR
fluxes can be seen in Figure \ref{fig:versus}. The best-fit power-law slope of 0.66$\pm$0.02 
is
consistent with that found for the large sample of NSXB, which did not
include \src, and also consistent with the expected slope from X-ray
heating  \citep{rufejo2007}. A viscously-heated disk would have
resulted in a slope of $\sim$0.3. \citet{rufejo2007} combined the NSXB
data regardless of spectral state. Dividing our data based on the
dominant X-ray spectral component (thermal vs. non-thermal) reveals a
significantly steeper relation for the (hard) non-thermal X-ray state
than for the (soft) thermal X-ray state (see Figure \ref{fig:versus}):
1.33$\pm$0.11 compared to 0.37$\pm$0.03. Although these different
slopes are obtained from small numbers of points, they are both no
longer consistent with X-ray heating. The slope for the (soft) thermal state
is marginally consistent with viscous heating, while the slope for the
(hard) non-thermal state is similar to that found between radio flux and
X-ray flux in NSXBs, $L_{radio} \propto {L_X}^{1.4}$ \citep{mife2006}. For this slope also to be observed in the nIR, the nIR-radio spectral index
would have to be very flat, requiring the nIR to originate from the optically-thick part of a jet synchrotron spectrum, for which we see no evidence however. 

An additional test to distinguish between X-ray and viscous heating is
to search for lags between changes in the nIR and X-ray emission of a
few days or more. Such lags are not expected when X-ray heating is
important, as reprocessing of the X-rays is near-instantaneous. If, on
the other hand, X-ray heating is not important, variations in the X-rays from the inner
disk will lag variations in the nIR emission from the non-irradiated outer disk, since changes in 
the mass-accretion rate travel through the disk on the viscous time scale.
Such delays have been observed  in GX 339-4 for example
\citep{hobuma2005}, where a lag time of 15--20 days was measured. In
Figure \ref{fig:lags} we plot the broadband X-ray flux of the thermal
component (disk and boundary layer; filled dots), the non-thermal
power-law component (open squares), and the $K_s$ flux density light
curves from an X-ray flux cycle in 2007 July-September that was sampled
well in the nIR. The result is ambiguous, mainly because the source
underwent a spectral transition in X-rays, making it hard to follow
the disk component over a long period of time; changes in the disk
X-ray flux during a transition probably do not only reflect changes in the
mass-accretion rate through the inner disk. However, the figure does
show that the nIR light curve resembles that of the total X-ray flux
(thermal and non-thermal combined), suggesting the nIR flux is produced by
X-ray heating. Better sampled light curves would be needed to perform
a more detailed study of time lags.

Based on
observed spectral indices \citet{rufejo2007} concluded that in atoll NSXBs with 2--10 keV luminosities
above $10^{36}$ erg\,s$^{-1}$ jet emission dominates the nIR. Distance estimates from  radius expansion bursts \citep[7.2--9.6 kpc][]{jone2004,gamuha2006}, indicate that
the 2--10 keV luminosity of \src\ is well above this value. However, our spectral
indices are more consistent with disk emission. The observed 0.5--100 keV flux range of \src\ leads to a total
luminosity range of $0.7-14.9\times10^{37}$ erg\,s$^{-1}$ (7.2 kpc) or
$1.2-26.5\times10^{37}$ (9.6 kpc). These values are quite high for atoll sources and come close to those of the Z sources, possibly indicating a larger accretion disk than is typical for atoll sources. This could then explain why X-ray heating dominates the nIR in \src.

We conclude that
X-ray heating of the outer disk and secondary star is the most likely source of the nIR emission in \src. Detection of possible orbital modulations and obtaining a nIR spectrum might help in distinguishing the relative importance of these two mechanisms.
More observations are needed to study a possible X-ray state
dependence of the nIR/X-ray relation. 

\acknowledgements We thank the numerous observers who have helped us
obtain data: F.~Baganoff, J.A.~Blackburne, A.~Burgasser, H.-W.~Chen,
D.~Erb, S.~Laycock, S.~Rappaport, P.~Schechter, and  M.~Torres. Partial support for DLK was 
provided by NASA through Hubble Fellowship grant
\#01207.01-A awarded by the Space Telescope Science Institute, which is
operated by the Association of Universities for Research in Astronomy, Inc.,
for NASA, under contract NAS 5-26555. PyRAF is a product of the Space Telescope Science 
Institute, which is
operated by AURA for NASA. This research has made use of data obtained from the High Energy Astrophysics Science Archive Research Center (HEASARC), provided by NASA's Goddard Space Flight Center.

{\it Facilities:} \facility{Magellan:Baade (PANIC)},
\facility{CTIO:Blanco (ISPI)}


\begin{deluxetable}{l l c c c c l  c c c c l c c c c}
\tablecaption{Near-IR Observation Summary for \src\label{tab:obs}}
\tablewidth{0pt}
\setlength{\tabcolsep}{3pt}
\tabletypesize{\scriptsize}
\tablehead{
\colhead{Epoch} & \colhead{Date\tablenotemark{\dagger}} & \multicolumn{4}{c}
{\ks} && \multicolumn{4}{c}{$H$}
&& \multicolumn{4}{c}{$J$} \\ \cline{3-6} \cline{8-11} \cline{13-16}
 & &\colhead{Time} & \colhead{Exp. (s)} & \colhead{Airmass} &\colhead{Seeing} && 
\colhead{Time} & \colhead{Exp. (s)} & \colhead{Airmass} &\colhead{Seeing} && 
\colhead{Time} & \colhead{Exp. (s)} & \colhead{Airmass} &\colhead{Seeing} \\
}
\startdata
1 \dotfill& 2006-May-16 & 06:06&  50& 1.04& 0.6&& 06:22  &  240& 1.04& 0.7 && 
06:11&  450& 1.04& 0.7\\
2 \dotfill& 2006-Jul-15 & 00:34& 230& 1.12& 0.7&& 00:47  & 1296& 1.10& 0.7 && 
01:19& 1620& 1.07& 0.7\\
3 \dotfill& 2006-Aug-04 & 00:45& 180& 1.04& 0.6&& 00:51  &  864& 1.04& 0.7 && 
01:18& 1620& 1.04& 0.7\\
4 \dotfill& 2006-Aug-08a & 00:45&  90& 1.04& 0.9&& 00:07  &  432& 1.05& 1.1 && 
00:20&  810& 1.04& 0.9\\
5 \dotfill& 2006-Aug-08b & 23:47& 180& 1.06& 1.0&& 23:54  &  432& 1.06& 1.0 && 
00:06& 1215& 1.05& 1.0\\
6 \dotfill& 2006-Aug-09 & 23:37&  90& 1.07& 0.7&& 23:43  &  432& 1.06& 0.8 && 
23:55&  810& 1.05& 0.9\\
7 \dotfill& 2006-Aug-26 & 23:30& 270& 1.04& 0.7&& 23:53  & 1296& 1.04& 0.7 && 
00:35& 2430& 1.06& 0.8\\
8 \dotfill& 2007-Apr-07 & 09:42& 100& 1.05& 0.6&& \nodata& \nod& \nod& \nod&& 
09:35&  100& 1.04& 0.6\\
9 \dotfill& 2007-May-09 & 07:56& 300& 1.06& 0.7&& \nodata& \nod& \nod& \nod&&  
\nod& \nod& \nod& \nod\\
10\dotfill & 2007-Jul-05& 01:08& 200& 1.16& 0.8&& \nodata& \nod& \nod& \nod&& 
01:31&  200& 1.12& 0.7\\
11\tablenotemark{\star}\dotfill & 2007-Jul-24& 01:53& 320& 1.03& 1.0&& \nodata& 
\nod& \nod& \nod&&  \nod& \nod& \nod& \nod\\
12\tablenotemark{\star}\dotfill & 2007-Jul-27& 06:31& 480& 1.96& 0.8&& \nodata& 
\nod& \nod& \nod&&  \nod& \nod& \nod& \nod\\
13\dotfill & 2007-Jul-31a & 01:45&  90& 1.04& 0.7&& 01:36  &  270& 1.04& 0.7 && 
01:27&  270& 1.04& 0.7\\
14\dotfill & 2007-Jul-31b & 23:25& 110& 1.16& 0.7&& \nodata& \nod& \nod& \nod&& 
23:34&  270& 1.14& 0.7\\
15\dotfill & 2007-Aug-03& 02:06& 300& 1.04& 0.6&& \nodata& \nod& \nod& \nod&&  
\nod& \nod& \nod& \nod\\
16\dotfill & 2007-Aug-20& 00:51& 180& 1.04& 0.6&& \nodata& \nod& \nod& \nod&& 
01:01&  360& 1.04& 0.5\\
17\dotfill & 2007-Sep-01& 02:46& 220& 1.30& 0.4&& \nodata& \nod& \nod& \nod&&  
\nod& \nod& \nod& \nod\\
18\dotfill & 2007-Sep-22& 01:38& 360& 1.35& 0.8&& 01:29  & 450 & 1.32&  0.7&& 
01:23&  180& 1.30&0.9\\
\enddata
\tablenotetext{\dagger}{All dates and times are UT. }
\tablenotetext{\star}{All observations were done with Baade/PANIC, except
  for these which used Blanco/ISPI.}
\end{deluxetable}

\begin{deluxetable}{l c c c c}
\tablecaption{Photometry of \src\label{tab:phot}}
\tablewidth{0pt}
\tablehead{\colhead{Epoch} & \colhead{X-ray} & \multicolumn{3}{c}{Magnitude} \\ 
\cline{3-5}
& state & \colhead{\ks} & \colhead{$H$} &
  \colhead{$J$} \\
}
\startdata
1  \dotfill& Soft &  $15.23\pm0.02$ & $16.03\pm  0.02 $ & $17.14\pm  0.02 $\\
2  \dotfill& ?\tablenotemark{\star} &  $16.18\pm0.02$ & $16.82\pm  0.02 $ & 
$17.96\pm  0.02 $\\
3  \dotfill& ?\tablenotemark{\star} &  $16.10\pm0.02$ & $16.93\pm  0.02 $ & 
$18.16\pm  0.02 $\\
4  \dotfill& Hard &  $16.98\pm0.08$ & $17.41\pm  0.03 $ & $18.47\pm  0.03 $\\
5  \dotfill& Hard &  $16.74\pm0.03$ & $17.37\pm  0.03 $ & $18.39\pm  0.03 $\\
6  \dotfill& Hard &  $16.83\pm0.03$ & $17.42\pm  0.03 $ & $18.48\pm  0.03 $\\
7  \dotfill& Hard &  $16.66\pm0.02$ & $17.34\pm  0.02 $ & $18.39\pm  0.02 $\\
8  \dotfill& Soft &  $15.51\pm0.02$ & \nodata		& $17.05\pm  0.02 $\\
9  \dotfill& Soft &  $15.77\pm0.02$ & \nodata		& \nodata	   \\
10 \dotfill& Soft &  $16.19\pm0.02$ & \nodata		& $17.94\pm  0.03 $\\
11 \dotfill& Hard &  $17.25\pm0.07$ & \nodata		& \nodata	   \\
12 \dotfill& Hard &  $17.10\pm0.06$ & \nodata		& \nodata	   \\
13 \dotfill& Hard &  $16.41\pm0.03$ & $16.94\pm  0.02 $ & $18.01\pm  0.02 $\\
14 \dotfill& Hard &  $16.54\pm0.03$ & \nodata		& $18.23\pm  0.03 $\\
15 \dotfill& Hard &  $16.31\pm0.02$ & \nodata		& \nodata	   \\
16 \dotfill& Soft &  $15.61\pm0.02$ & \nodata		& $17.10\pm  0.02 $\\
17 \dotfill& Soft &  $15.37\pm0.02$ & \nodata		& \nodata	   \\
18 \dotfill& Soft &  $15.60\pm0.02$ & $16.18\pm  0.02 $ & $17.14\pm  0.02 $\\
\enddata

\tablenotetext{\star}{The 
state identification of these two observations is uncertain.}
\tablecomments{The quoted errors only reflect \texttt{sextractor}
  uncertainties. Additional uncertainties related to flat fielding and sky subtraction are estimated 
to be $0.05$~mag.}
\end{deluxetable}


\newpage
\clearpage

\begin{figure} 
\centerline{\includegraphics[width=12cm]{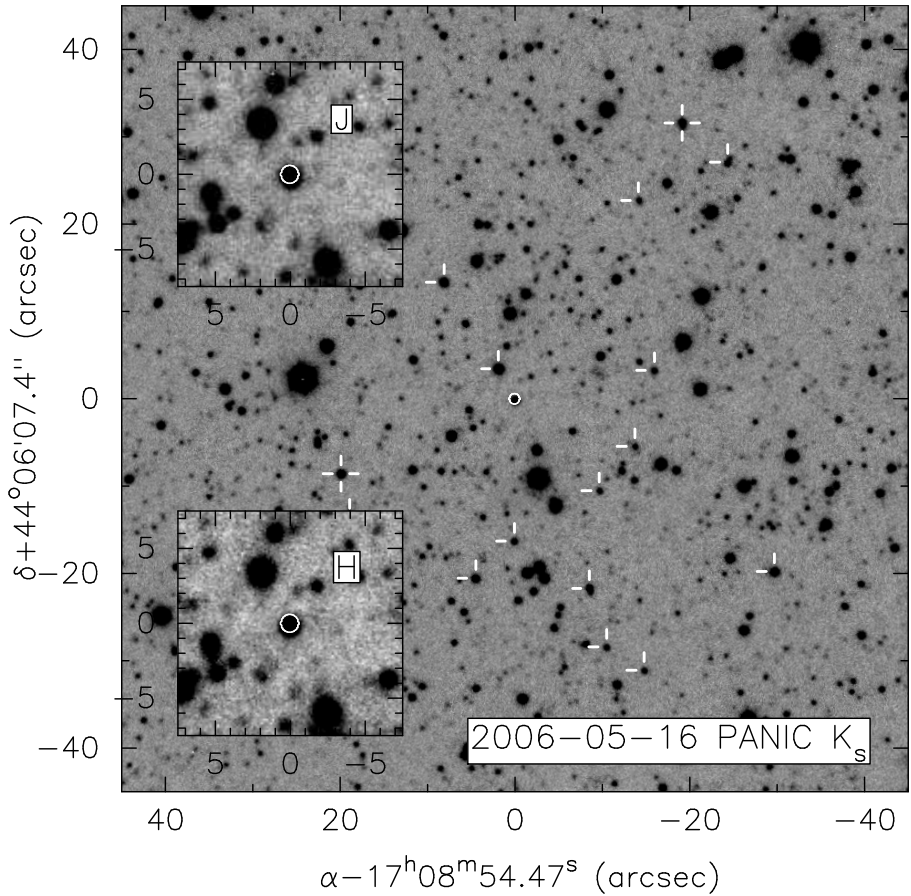}}
\caption{Images of the field around \src\ from the 2006~May~16 PANIC
observation.  The main panel shows a $1\farcm5\times1\farcm5$ field of the \ks\
image.  The \chandra\ position of \src\ is marked with a
$0\farcs6$-radius circle. We also indicate with tick marks a number of
the photometric reference stars (some are hidden behind the sub-panels
or are off this image): the two most stable stars have crosses, while the
rest have only ticks to the north and east.   The sub-panels show
$15\arcsec\times15\arcsec$ fields of the $J$ (top left) and $H$ (bottom left)
images.  North is up, and east to the left.  } \label{fig:jhk}
\end{figure}


\begin{figure}
\centerline{\includegraphics[width=12cm]{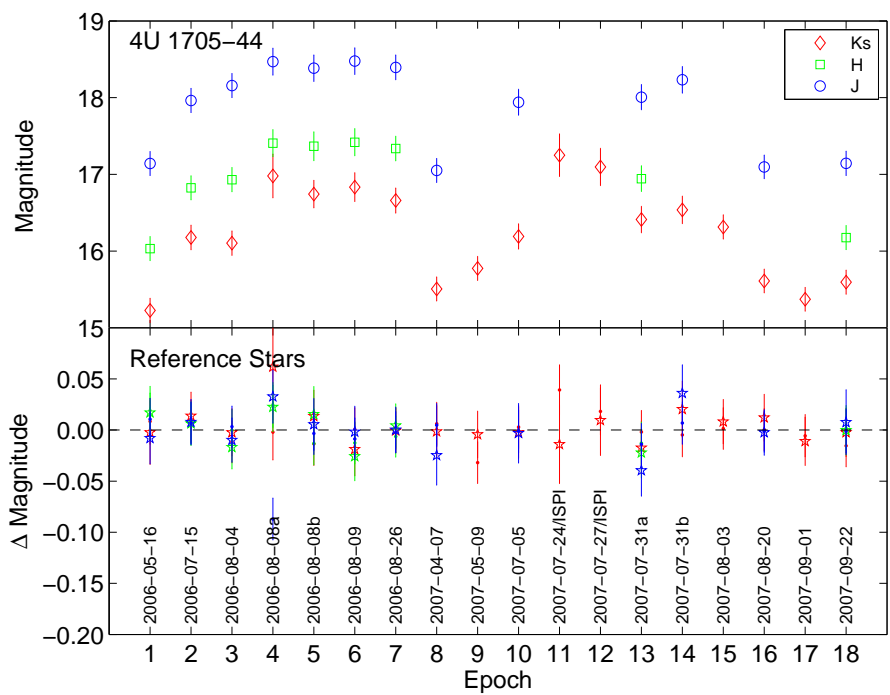}}
\caption{Photometry of \src.  In the top panel we show the \ks\ (diamonds), $H$ (squares), and $J$ (circles) photometry of
  \src\ plotted as a function of epoch number.  In the bottom panel we
  show the variation in the magnitudes of the two most stable reference stars
  (see Fig.~\ref{fig:jhk}).  The colors/gray-scales represent the same filters, but here the two symbols (points and stars) correspond to the two
  reference stars.  The date of each epoch is listed in the bottom panel.  [{\it(See the electronic edition of the Journal for a color version of this figure.}] }
\label{fig:phot}
\end{figure}

\begin{figure}
\centerline{\includegraphics[angle=-90,width=8cm]{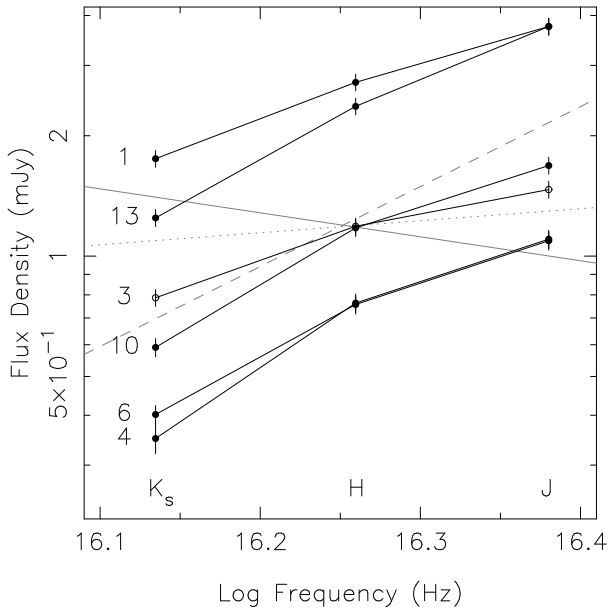}}
\caption{Six representative nIR spectral energy distributions (SEDs)
of \src, de-reddened assuming an $N_H$ of 1.9$\times10^{22}$
atoms\,cm$^{-2}$ ($A_V=10.3$). The nIR epoch of each of the SEDs is
shown on the left-side of the $K_s$ data points. The solid grey line
depicts a typical optically thin synchrotron spectrum from a jet
($\propto\nu^{-0.6}$).  The dashed line represents the Rayleigh-Jeans
tail of the spectrum of a viscously heated accretion disk
($\propto\nu^{2}$) and the dotted line the shallower 'stretched' part
that is characteristic of such a disk spectrum ($\propto\nu^{1/3}$).  }
\label{fig:sed} \end{figure}

\begin{figure}
\centerline{\includegraphics[angle=-90,width=11cm]{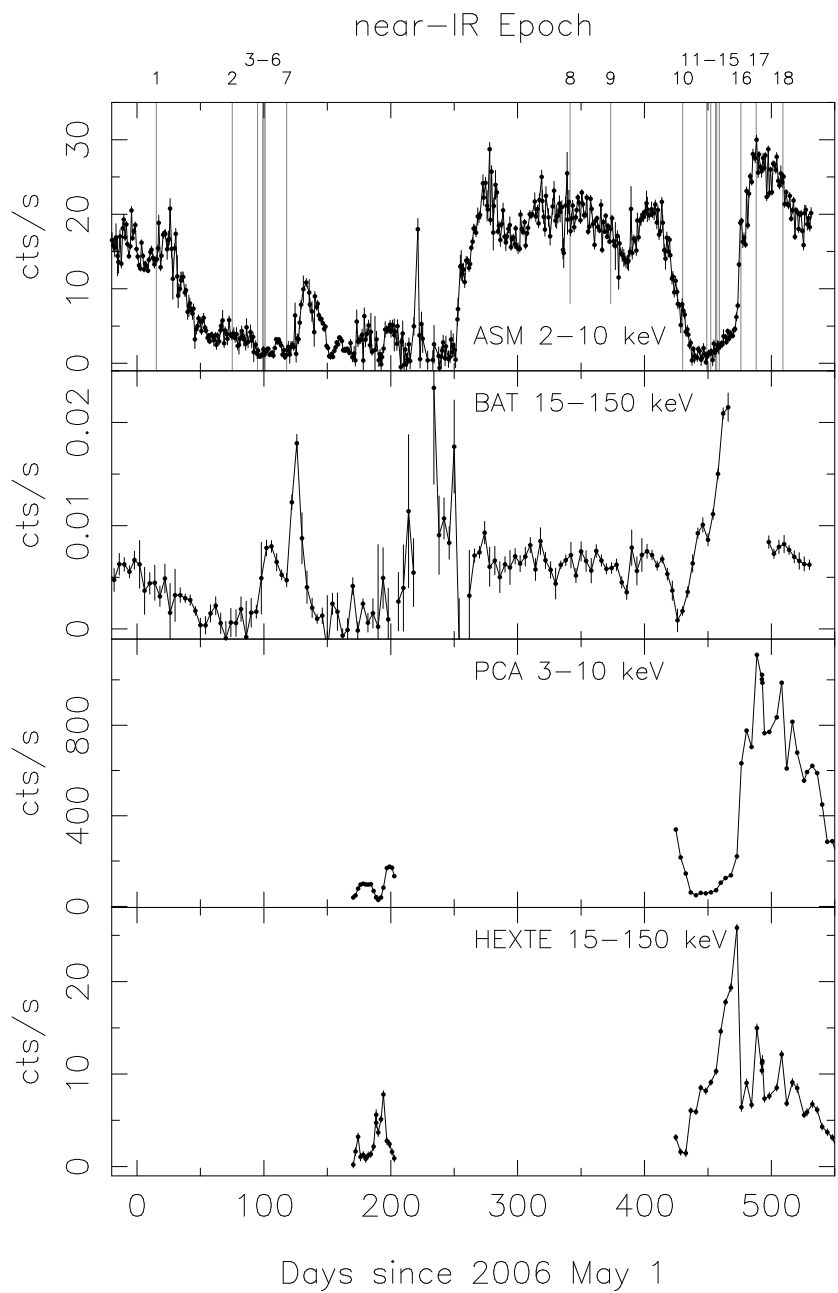}}
\caption{X-ray light curves of \src\ from various instruments. From
top to bottom: {\it RXTE}/ASM, {\it Swift}/BAT, {\it RXTE}/PCA, and
{\it RXTE}/HEXTE. The ASM and BAT light curves show one- and four-day
averages, respectively, whereas the PCA and HEXTE light curves were
rebinned to one data point per observation. The times of the nIR
observations are indicated by the vertical lines and numbers in and
above the top panel.} \label{fig:curve} \end{figure}

\begin{figure}
\centerline{\includegraphics[angle=-90,width=10cm]{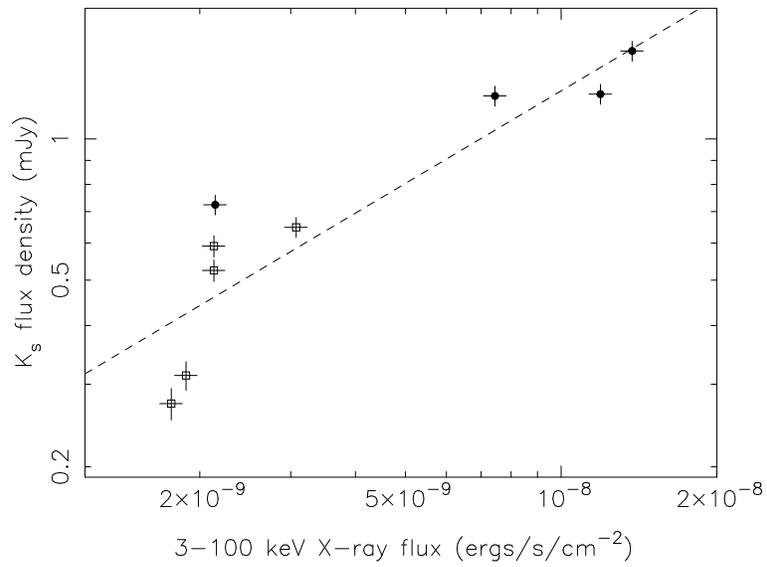}}
\caption{$K_s$ flux density versus unabsorbed X-ray flux (3--100 keV). The dashed
line shows the best power-law fit to the data with an index of 0.66(2).
Filled circles correspond to observations in which the thermal
components dominates the X-ray spectrum ($>85\%$), open squares ones in which
the non-thermal flux dominates ($>75\%$).} \label{fig:versus} \end{figure}

\newpage
\clearpage

\begin{figure}
\centerline{\includegraphics[angle=-90,width=12cm]{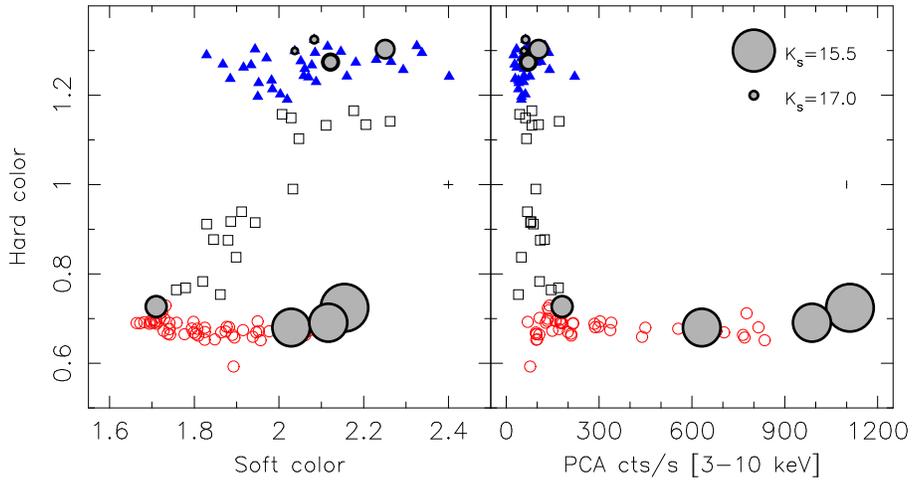}}
\caption{ Color-color (left) and color-count rate diagrams (right) of
\src, created from {\it RXTE}/PCA data. Soft color and hard color were
defined as count rate ratios in the bands (4.0--7.3 keV)/(2.3--4.0
keV) and (9.8--18.2 keV)/(7.3--9.8 keV), respectively. Typical error
bars are shown on the right hand side of each panel. Triangles, squares, and
open circles indicate the spectrally hard (``extreme island"),
transitional (``island"), and soft (``banana") states, respectively. The gray circles
depict observations coincident (within two days) with our nIR observations, with the size of the circles scaling linearly
with nIR flux density.  [{\it(See the electronic edition of the Journal for a color version of this figure.}] } \label{fig:cd} \end{figure}

\begin{figure}
\centerline{\includegraphics[angle=-90,width=12cm]{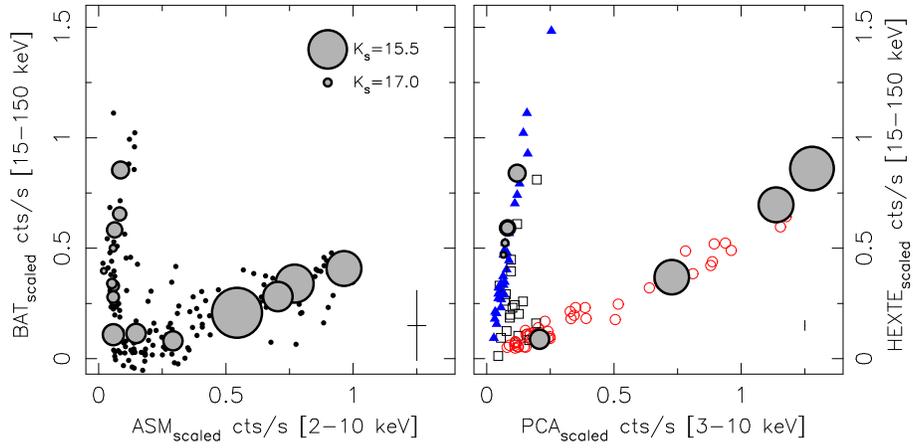}}
\caption{High-energy versus low-energy scaled count rates from
ASM/BAT (left) and PCA/HEXTE (right). The count rates were scaled as
described in \S\ref{sec:correlations}. Typical error bars are shown in the 
lower right corner of each panel. Triangles, squares, and
open circles in the right panel indicate
the spectrally hard (``extreme island"), transitional (``island"), and
soft (``banana") states, respectively. The gray circles depict observations coincident (within two days) with our nIR observations, with
the size of the circles scaling linearly with nIR flux density. [{\it(See the electronic edition of the Journal for a color version of this figure.}] }
\label{fig:iid} \end{figure}

\newpage
\clearpage


\begin{figure}
\centerline{\includegraphics[angle=-90,width=11cm]{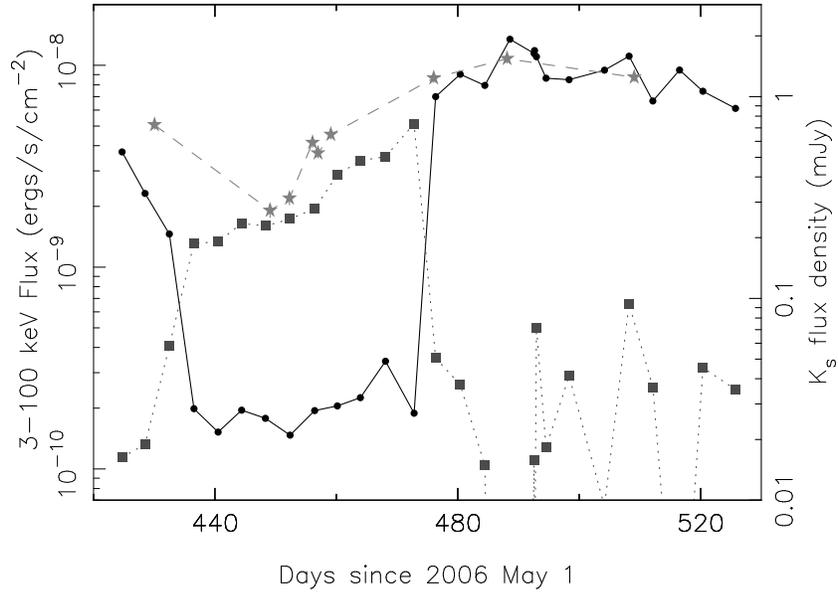}}
\caption{Comparison of thermal (dots) and non-thermal (squares) X-ray
light curves (3--100 keV) and $K_s$-band (stars) light
curves during a transition cycle in July-September 2007.}
\label{fig:lags} \end{figure}

\end{document}